\documentclass[aps,pre,twocolumn,superscriptaddress,showpacs]{revtex4}
\usepackage{amssymb}
\usepackage{epsfig}
\usepackage{amsmath}
\usepackage{times}

\begin{document}

\title{A model bridging chimera state and explosive synchronization }

\author{Xiyun Zhang}
\affiliation{Department of Physics, East China Normal University,
Shanghai, 200062, P. R. China}

\author{Hongjie Bi}
\affiliation{Department of Physics, East China Normal University,
Shanghai, 200062, P. R. China}

\author{Shuguang Guan}
\affiliation{Department of Physics, East China Normal University,
Shanghai, 200062, P. R. China}

\author{Jinming Liu}
\affiliation{State Key Laboratory of Precision Spectroscopy,
East China Normal University, Shanghai 200062, China}

\author{Zonghua Liu}
\email{zhliu@phy.ecnu.edu.cn} \affiliation{Department of Physics,
East China Normal University, Shanghai, 200062, P. R. China}
\affiliation{State Key Laboratory of Precision Spectroscopy,
East China Normal University, Shanghai 200062, China}

\date{\today}

\begin{abstract}
Global and partial synchronization are the two distinctive forms of synchronization in coupled oscillators
and have been well studied in the past decades. Recent attention on synchronization is focused on the chimera
state (CS) and explosive synchronization (ES), but little attention has been paid to their
relationship. We here study this topic by presenting a model to bridge these two phenomena, which consists
of two groups of coupled oscillators and its coupling strength is adaptively controlled by a local order
parameter. We find that this model displays either CS or ES in two limits. In between
the two limits, this model exhibits both CS and ES, where CS can be observed for a fixed coupling strength
and ES appears when the coupling is increased adiabatically. Moreover, we show both theoretically and
numerically that there are a variety of CS basin patterns for the case of identical oscillators, depending on
the distributions of both the initial order parameters and the initial average phases. This model suggests a
way to easily observe CS, in contrast to others models having some (weak or strong) dependence on initial
conditions.

\end{abstract}


\maketitle

\section{Introduction}
Synchronization in coupled oscillators has been well studied in the past decades and is now focused on the
influence of network structures \cite{Pikvosky:2003,Boccaletti:2006,Arenas:2008}. In this field, two hot
topics are the chimera state (CS) and explosive synchronization (ES), respectively. CS was first found by
Kuramoto and Battogtokh in 2002 \cite{Kuramoto:2002}.  After the discovery, CS has attracted
a lot of attention in the past decade \cite{Omel:2008,Sethia:2008,Bordyugov:2010,Laing:2009,Martens:2009,Wolfrum:2011,
Laing:2012,Zhu:2012,Panaggio:2013,Dudkowski:2014,Omelchenko:2015,Jaros:2015,Bohm:2015}. Generally speaking, CS
is the coexistence of coherent and incoherent behaviors in coupled identical oscillators. Because of different
initial conditions, the nonlocally coupled oscillators naturally evolve into distinct coherent and
incoherent groups. This counterintuitive coexistence of coherent and incoherent oscillations in populations
of identical oscillators, each with an equivalent coupling structure, can be considered as a symmetry breaking
on the collective behavior by nonsymmetric initial conditions. This phenomenon reminded people the two heads
monster in Greek mythology and thus was named as {\sl Chimera State} by Abrams and Strogatz in 2004
\cite{Abrams:2004}.  The study of CS was originally motivated by the phenomenon of unihemispheric sleep of many
creatures in real world \cite{Rattenborg:2000,Mathews: 2006,Abrams:2008,Pikvosky:2008,Ma:2010},
which was first found in dolphin and then revealed in birds, some aquatic mammals, and reptiles etc. So far,
CS has been confirmed in many experiments
\cite{Tinsley:2012,Hagerstrom:2012,Viktorov:2014,Wickramasinghe:2014,Martens:2013,Larger:2013,Schoenleber:2014}.
For example, Tinsley {\sl et al} reported on experimental studies of CS in populations of coupled chemical
oscillators \cite{Tinsley:2012}. Hagerstrom {\sl et al} showed experimental observation of CS in
coupled-map lattices \cite{Hagerstrom:2012}. Viktorov {\sl et al} demonstrated a coexistence of coherent and
incoherent modes in the optical comb generated by a passively mode-locked quantum dot laser \cite{Viktorov:2014}.
Wickramasinghe {\sl et al} presented the experiment of CS in a network of electrochemical reactions
\cite{Wickramasinghe:2014}. Martens {\sl et al} devised a simple experiment with mechanical oscillators to
show CS \cite{Martens:2013}. And Schoenleber {\sl et al} reported the CS in the oxide layer during the oscillatory
photoelectrodissolution of n-type doped silicon electrodes under limited illumination \cite{Schoenleber:2014}.

ES represents the first-order synchronization transition in networked oscillators. When we increase the coupling
strength adiabatically, the system stays unsynchronized until a critical forward coupling strength $\lambda_{cF}$
where the system suddenly becomes synchronized. That is, its order parameter $R$ has a jump at $\lambda_{cF}$.
However, when we decrease the coupling strength adiabatically from a synchronized state, the system does not go
back by the same route as the forward process but jump at a different critical backward coupling strength
$\lambda_{cB}$. As $\lambda_{cF}>\lambda_{cB}$, the forward and backward routes of $R$ forms a hysteresis loop.
This first-order transition was
in fact found before the concept of complex networks \cite{Strogatz:1989,Tanaka:1997,Pazo:2005} and became hot
only when it was rediscovered from the positive correlation between the natural frequency of a networked oscillator
and its degree by G\'{o}mez-Garde\~{n}es {\sl et al} and named as {\sl Explosive Synchronization} in 2011
\cite{Gomez:2011}. Before the work \cite{Gomez:2011}, synchronization on complex networks was generally analyzed
by the approach of master stability function \cite{Pecora:1998}, which always predicts a second-order phase transition.
However, the work \cite{Gomez:2011} showed that it is also possible
for the synchronization on complex networks to be the first-order, thus inducing great attention on ES \cite{Leyva:2012,Peron:2012,Coutinho:2013,Liu:2013,Ji:2013,Zhang:2013,Leyva:2013,Leyva:2014,Su:2013,Hu:2014,
Zou:2014,Zhou:2015,Zhang:2015}. It was revealed that except the way in \cite{Gomez:2011}, ES can be also observed
by many other ways, providing that the growth of synchronized clusters is under a suppressive rule
\cite{Zhang:2014}.

Currently, CS and ES are separately studied as two distinctive topics. In general, we do not have CS in the systems
of ES, and vice versa. Thus, it is interesting to ask whether it is possible to observe both of them in a single
system. To figure out the answer, we here study this topic by presenting a novel model to bridge these two phenomena.
The model consists
of two groups of coupled nonidentical oscillators with a natural frequency distribution. Specifically, its coupling
strength is adaptively controlled by a parameter $\beta$. This model goes back to the standard CS
model \cite{Abrams:2008} when all the natural frequencies are the same and $\beta=0$ and returns to the adaptive
model of ES \cite{Zhang:2015} when there is only one group of oscillators and $\beta=1$.
 Very interesting, we find that this model displays both CS and ES, where CS can be observed for a fixed
coupling strength and ES appears when the coupling is increased adiabatically. Thus, this model sets up a bridge
between CS and ES. Moreover, we focus on the case of identical oscillators and show both theoretically and numerically
that there are a variety of CS basin patterns, depending on the distributions of both the initial order parameters
and the initial average phases. That is, this model shows a way to easily observe CS, in contrast to the sensitive
dependence on initial conditions in many previous models \cite{Martens:2016,Feng:2015}.

The paper is organized as follows. In Sec.II, we introduce the model and study its collective behaviors. In Sec.III,
we pay attention to the case of identical oscillators and study it by the dimensional reduction analysis. In Sec. IV,
we show the corresponding numerical simulations and and its stability analysis. Finally, in Sec. V, we give
conclusions and discussions.

\section{Model description}
We consider a model of two groups of coupled oscillators, defined as
\begin{eqnarray}
\label{model}
\dot{\theta}_{i,j}&=&\omega_{i,j}+\frac{R_j^{\beta}\lambda}{N}\sum_{k=1}^{N}\sin(\theta_{k,j}-\theta_{i,j}+\alpha) \nonumber\\
& & +\frac{R_j^{\beta}\lambda'}{N}\sum_{k=1}^{N}\sin(\theta_{k,j'}-\theta_{i,j}+\alpha),
\end{eqnarray}
where the index $j=1,2$ represents the two groups, $i=1,\cdots,N$ represents the $N$ oscillators in each group.
$\omega_{i,j}$ is the natural frequency satisfying an uniform distribution in $(-\delta,\delta)$. The oscillators
are globally coupled with coupling strength $\lambda$ inside each group and coupling strength $\lambda'$ between the
two groups. $j'$ represents the other group, defined as $j'=2$ when $j=1$ and $j'=1$ when $j=2$. $\alpha$ is a phase
lag parameter and set as $\alpha=\frac{\pi}{2}-0.1$, which was chosen by many CS papers
\cite{Abrams:2008,Pikvosky:2008,Ma:2010}. The coupling is attractive when $\alpha<\pi/2$ and repulsive when
$\alpha>\pi/2$. $\beta$ is a parameter located in $[0,1]$.

\begin{figure}
\epsfig{figure=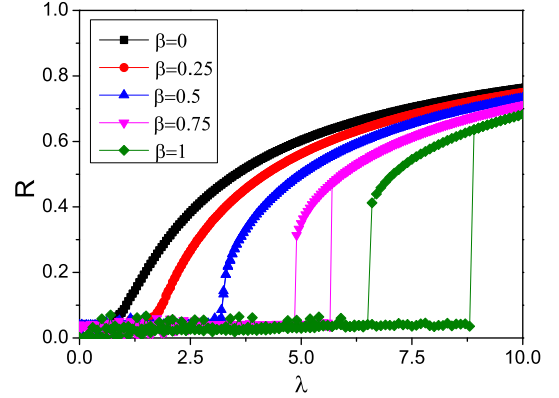,width=1.\linewidth} \caption{(color online).
Synchronization transition of model (\ref{model}) with only one population, i.e. $\lambda'=0$. The parameters are
$\delta=1.0$ and $\alpha=\frac{\pi}{2}-0.1$}
\label{Fig:model}
\end{figure}

$R_1$ and $R_2$ in Eq. (\ref{model}) are the order parameters of the groups $1$ and $2$, respectively, which are
defined as
\begin{equation}
\label{order}
R_1e^{i\Psi_1}=\frac{1}{N}\sum_{k=1}^N e^{i\theta_{k,1}}, \quad
R_2e^{i\Psi_2}=\frac{1}{N}\sum_{k=1}^N e^{i\theta_{k,2}}.
\end{equation}
In the framework of Eq. (\ref{model}), the population is put into two groups and the coupling strengthes
$R_j^{\beta}\lambda$ and
$R_j^{\beta}\lambda'$ are closely correlated to the local coherence when $\beta$ is not $0$. The model (\ref{model})
will return to the case of one population in Ref. \cite{Zhang:2015} when $\lambda'=0$ and $\beta=1$. To show the
influence of $\beta$, Fig. \ref{Fig:model} shows the synchronization transition of model (\ref{model}) for different
$\beta$, with $\lambda'=0$. It's easy to see that $R$ has a continuous transition for $\beta=0$, a discontinuous
transition for $\beta=1$, and a transition gradually changing from continuous to discontinuous when $\beta$ increases,
indicating a transition from traditional synchronization to explosive synchronization. When $\beta$ is in the range
of hysteresis loop, there are bistability where the final state of system depends sensitively on the initial
conditions.

The model (\ref{model}) is more sensitive to the local coherence if there are two or more groups in the system. Once
the initial conditions are asymmetric, the two groups may easily go to different final states, i.e. one group with high
coherence and another group with low coherence.

\begin{figure}
\epsfig{figure=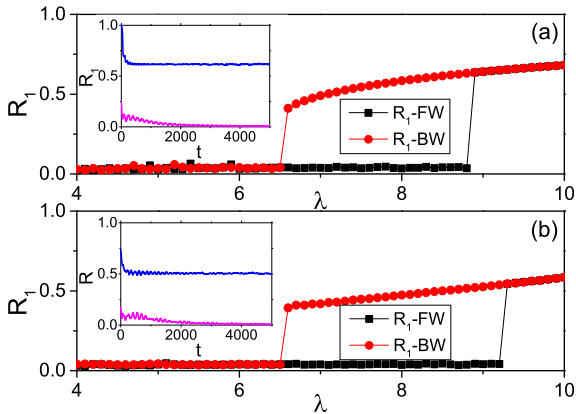,width=1.\linewidth} \caption{(color online).
(a) Case of $\lambda'=0$ with $\delta=1.0$, $\alpha=\frac{\pi}{2}-0.1$ and $\beta=1$. The ``squares" and
``circles" represent
$R_1$ for the forward and backward continuation diagram, respectively. The inset shows the evolution of two different
initial conditions for $\lambda=8.5$. (b) Case of $\lambda'=2$ with $\delta=1.0$, $\alpha=\frac{\pi}{2}-0.1$ and
$\beta=1$. The ``squares" and ``circles" represent $R_1$ for the forward and backward continuation diagram
($R_2$ has the same loop but not shown here), respectively. The two curves in the inset show the
evolution of two typical $R_1$ and
$R_2$ in the two groups, respectively, with $\lambda=8.5$.}
\label{Fig:bistability}
\end{figure}

Eq. (\ref{model}) has two limiting behaviors. The first one is the limiting behavior of $\lambda'=0$ and $\beta=1$,
which goes back to the
adaptive model of ES in Ref. \cite{Zhang:2015}. In this situation, ES can be observed if we increase (decrease) the
coupling adiabatically in the forward (backward) continuation diagram. Fig. \ref{Fig:bistability}(a) shows the dependence
of $R_1$ on $\lambda$ for $\delta=1.0$. It is easy to see that there is a hysteresis loop, indicating the existence
of ES. The inset
of Fig. \ref{Fig:bistability}(a) shows the evolution of two different initial conditions for $\lambda=8.5$. We see
that one gradually approaches a higher value ($R_1\approx 0.58$) and the other goes to zero, confirming the
sensitivity to initial conditions in the bistable region. We have the same results for $R_2$ of another group
(it's not shown in Fig. \ref{Fig:bistability} ),
as the system exhibits the symmetry $1\leftrightarrow 2$.

\begin{figure}
\epsfig{figure=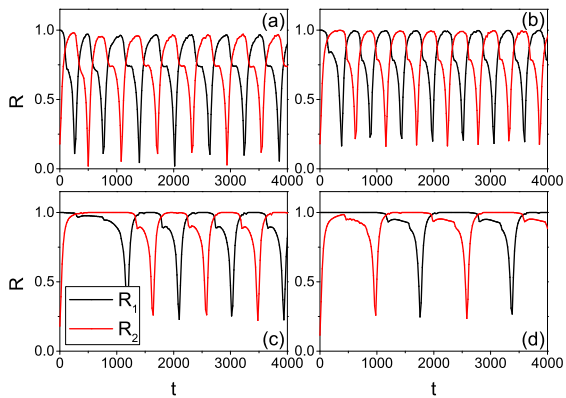,width=1.\linewidth} \caption{(color online).
The chimera-like behaviors in Eq. (\ref{model}) for $\beta=0, \lambda=8.0$ and $\lambda'=3.0$ where the
black and red line represent $R_1$ and $R_2$, respectively, and (a)-(d)
represent the cases of $\delta=1.0, 0.5, 0.2$ and $0.15$, respectively.}
\label{Fig:chimera-like}
\end{figure}

The second one is the limiting behavior of identical $\omega_{i,j}$ ($\delta=0$) in Eq. (\ref{model}) for all the
oscillators and
$\beta=0$, which returns to the typical model of CS in Ref. \cite{Abrams:2008}. In this case, our numerical
simulations confirm that one group is synchronized with $R_1=1$ while the other is unsynchronized with
$R_2<1$. Furthermore, we were surprised to find that there is still a chimera-like behavior when we keep $\beta=0$
but let $\omega_{i,j}$ satisfy the uniform distribution in $(-\delta,\delta)$. Figure \ref{Fig:chimera-like}
(a)-(d) show the results for $\delta=1.0, 0.5, 0.2$ and $0.15$, respectively. We see that the oscillation
periods of $R_1$ and $R_2$ increase with the decrease of $\delta$ until $\delta=0.15$. After that, the
oscillation behaviors of $R_1$ and $R_2$ will disappear and are replaced by one group synchronized and the other
unsynchronized, i.e. chimera state.

We now go back to the current model of Eq. (\ref{model}) with $\beta=1$. We find that it
can also show the hysteresis loop. Fig. \ref{Fig:bistability}(b) shows the results of $R_1$ for $\lambda'=2$.
Comparing Fig. \ref{Fig:bistability}(b) with Fig. \ref{Fig:bistability}(a) we see that their forward jumping
positions are slightly different, i.e. $\lambda_{cF}<9.0$ in Fig. \ref{Fig:bistability}(a) while $\lambda_{cF}>9.0$
in Fig. \ref{Fig:bistability}(b). We have observed the same results for $R_2$ (not shown here), as the symmetry
$1\leftrightarrow 2$ in the two groups of the system. The inset of Fig. \ref{Fig:bistability}(b) shows the
evolution of two typical initial conditions from the two groups, respectively, for $\lambda=8.5$. We see that
one ($R_1$) goes to a higher value ($R_1\approx 0.58$) and the other ($R_2$) goes to zero, indicating a
chimera-like behavior. Therefore, we have observed both ES and CS in the model of Eq. (\ref{model}) when the
parameters are taken in the range of the hysteresis loop.

\begin{figure}
\epsfig{figure=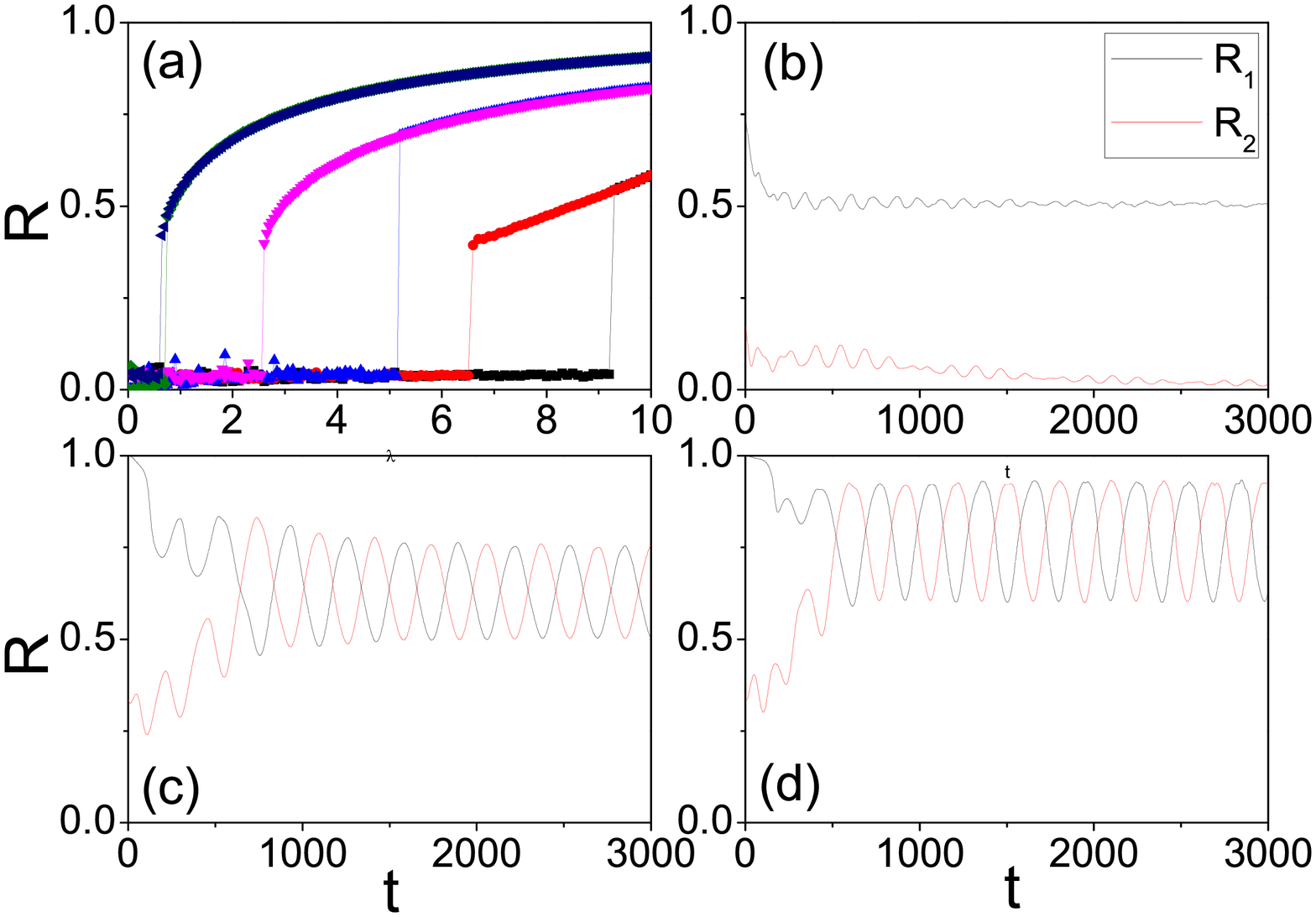,width=1.\linewidth} \caption{(color online).
Coexistence of ES and CS in the model of Eq. (\ref{model}) for $\alpha=\frac{\pi}{2}-0.1, \beta=1$, and
$\lambda'=2$. (a) $R_1$ versus $\lambda$ where the ``black"
and ``red" lines represent $R_1$ of the forward and backward continuation diagram for $\delta=1$, respectively;
the ``blue" and ``pink" lines represent the case of $\delta=0.7$; and the ``dark green" and
``dark blue" lines represent the case of $\delta=0.4$. (b)-(d) Evolutions of $R_1$ and $R_2$ on time $t$
for $\lambda=8.0$ and $\delta=1, 0.7$ and $0.4$, respectively.}
\label{Fig:chimera-hysteresis}
\end{figure}

Then, we change the range of frequency distribution $\delta$. We find that the hysteresis loop
depends on the parameter $\delta$ and can be observed only when $\delta>0.31$. With the decrease of $\delta$,
the size of the loop decreases until zero at about $\delta=0.31$ and the transition points of $R_1$ or
$R_2$ also approaches zero. Fig. \ref{Fig:chimera-hysteresis}(a) shows the results for $\lambda'=2$, where the
``squares" and ``circles" represent $R_1$ of the forward and backward continuation diagram for $\delta=1$,
respectively; the ``up triangles" and ``down triangles" represent the case of $\delta=0.7$; and the
``diamonds" and ``left triangles" represent the case of $\delta=0.4$. To check the coexistence of CS,
we study the evolution of $R_1$ and $R_2$ for two different initial conditions.
Fig. \ref{Fig:chimera-hysteresis}(b)-(d) show the results for $\lambda=8.0$ and $\delta=1, 0.7$ and $0.4$,
respectively. We see that Fig. \ref{Fig:chimera-hysteresis}(b) is a chimera-like state while
Fig. \ref{Fig:chimera-hysteresis}(c) and (d) are breather-like states. In sum, the range of frequency
distribution $\delta$ takes a key role for the coexistence of ES and CS.

\section{Dimensional reduction analysis}
In the following parts, we study CS in model (\ref{model}) with $\beta=1$. In order to satisfy the definition of CS,
we change to identical oscillators ($\delta=0$).
To make a theoretical analysis on Eq. (\ref{model}), it is better to reduce its dimension. Fortunately,
such an approach of dimensional reduction has been proposed by Watanabe and Strogatz \cite{Watanabe:1994} and
then generalized by Pikovsky and Rosenblum \cite{Pikovsky:2008}. We here adopted it to analyze the model
(\ref{model}). In a mean-field framework, the coupling terms in Eq. (\ref{model}) can be
rewritten as $R_a^2\lambda\sin(\Psi_a-\theta_{j}^a+\alpha)+R_aR_{a'}\lambda'\sin(\Psi_{a'}-\theta_j^a+\alpha)$.
Thus, Eq. (\ref{model}) can be rewritten as
\begin{eqnarray}
\label{meanfield}
\dot{\theta}_{j}^{a}&=&Im(Z_ae^{-i\theta_j^a}) \nonumber\\
Z_a&=&R_a^2\lambda e^{-i(\Psi_a+\alpha)}+R_aR_{a'}\lambda' e^{-i(\Psi_{a'}+\alpha)},
\end{eqnarray}
where $Z$ is the mean field coupling for the oscillator $j$, $a$ and $a'$ are the index of the two populations,
respectively. The average frequency $\langle \omega \rangle$ has been ignored as it is zero for a symmetric
distribution.  By introducing three variables $\rho_a(t)$, $\Theta_a(t)$, $\Phi_a(t)$ and constants $\psi_j^a$
via the transformation
\begin{equation}
\tan[\frac{\theta_j^a-\Phi_a}{2}]=\frac{1-\rho_a}{1+\rho_a}\tan[\frac{\psi_j^a-\Theta_a}{2}],
\end{equation}
we get the WS equations of the Eq. (\ref{model}) \cite{Watanabe:1994,Pikovsky:2008}
\begin{eqnarray}
\label{WSequation}
\dot{\rho}_a&=&\frac{1-\rho_a^2}{2}Re(Z_ae^{-i\Phi_a}) \nonumber\\
\dot{\Theta}_a&=&\frac{1-\rho_a^2}{2\rho_a}Im(Z_ae^{-i\Phi_a}) \nonumber\\
\dot{\Phi}_a&=&\frac{1+\rho_a^2}{2\rho_a}Im(Z_ae^{-i\Phi_a}).
\end{eqnarray}
Generally, the parameter $\rho$ characterizes the degree of synchronization: $\rho=0$, if the oscillators are
incoherent, and $\rho=1$, if the oscillators are complete synchronized. $\rho_a$ is roughly proportional to the
order parameter $R_a$. The phase variable $\Theta$ describes the shift of individual oscillators with the mean
phase and $\Phi$ describes the average of the phases. It is convenient to introduce new variables
$\xi_a=\Phi_a-\Theta_a$ and $z_a=\rho_ae^{i\Phi_a}$, then Eq. (\ref{WSequation}) can be rewritten as
\begin{eqnarray}
\dot{z}_a&=&\frac{1}{2}Z_a-\frac{z_a^2}{2}Z_a^* \label{lowdimesion1}\\
\dot{\xi}_a&=&Im(z_a^*Z_a) \label{lowdimesion2}.
\end{eqnarray}
If the constants $\psi_j^a$ are uniformly distributed, Eqs. (\ref{lowdimesion1}) and (\ref{lowdimesion2})
will decouple. Eq. (\ref{lowdimesion1}) describes the low dimensional behavior of Eq. (\ref{model}). In the
thermodynamic limit, we have $\rho_a=R_a$ and thus from Eq. (\ref{lowdimesion1}) we obtain
\begin{eqnarray}
\label{lowdimesion3}
\dot{R}_a&=&\frac{1}{2}R_a(1-R_a^2)[\lambda R_a\cos{\alpha}+\lambda' R_{a'}\cos{(\Phi_{a'}-\Phi_a+\alpha)}] \nonumber\\
\dot{\Phi}_a&=&\frac{1}{2}(1+R_a^2)[\lambda R_a\sin{\alpha}+\lambda' R_{a'}\sin{(\Phi_{a'}-\Phi_a+\alpha)}].
\end{eqnarray}
Eq. (\ref{lowdimesion3}) describes the theoretical prediction of the collective behaviors of Eq. (\ref{model}).
However, it is not easy to get the precise solution of Eq. (\ref{lowdimesion3}). Hence, we here calculate
Eq. (\ref{model}) numerically. In this way, the initial order parameters $R_1(0)$ and $R_2(0)$ and the initial
phases $\Phi_1(0)$ and $\Phi_2(0)$ will be the key factors to influence the final states $R_1$ and $R_2$.

\section{Results and analysis}
\begin{figure}
\epsfig{figure=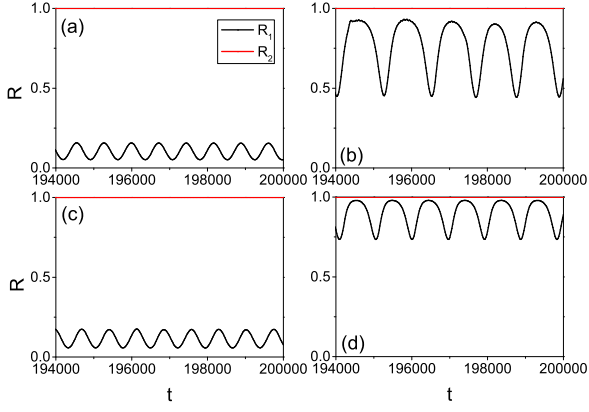,width=1.0\linewidth} \caption{(color
online). Comparison between numerical simulations and theoretical results where (a) and (b) represent the numerical
simulations from Eq. (\ref{model}) with $\omega_{i,j}=0$ and (c) and (d) the corresponding theoretical results
from Eq. (\ref{lowdimesion3}). (a) and (c): The
coupling strength is $\lambda=\lambda'=1$ and the initial order parameters are taken as $R_1(0)=0.275$ and
$R_2(0)=0.569$. The difference between the initial average phase is $\Delta\Phi=\Phi_2(0)-\Phi_1(0)=2\pi/3$.
(b) and (d): The coupling strengthes are $\lambda=1.5$ and $\lambda'=1$ and the initial order parameters are taken
as $R_1(0)=0.1$ and $R_2(0)=0.569$. The difference between the initial average phases is also taken as
$\Delta\Phi=\Phi_2(0)-\Phi_1(0)=2\pi/3$. }
\label{Fig:breath}
\end{figure}

In numerical simulations, we take the system size as $2N=100$, i.e. $N=50$ for each group. For the convenience
of comparing with the above theoretical predictions, we let all the natural frequencies $\omega_{i,j}$ in
Eq. (\ref{model}) be zero. The initial phases are drawn from the circular Cauchy distribution \cite{Mccullagh:1996}
\begin{equation}
g(\theta(0))=\frac{1-|\gamma|^2}{2\pi|e^{i\theta}-\gamma|^2}
\end{equation}
which can be easily generated from a Lorentzian distribution
$g(x)=\frac{1}{\pi}[\frac{\eta}{(x-x_0)^2+\eta^2}]$ with $\eta$ being the half width at half maximum and
$x_0$ being the center frequency. Making a transformation $X=\frac{x+i}{x-i}$, we can get a new complex variable
$X$, which is distributed on a unit circular in complex plane. The phases of $X$ are distributed as circular Cauchy
distribution. By changing $x_0$ and $\eta$, we can easily change the average and deviation of the circular Cauchy
distribution and thus change the initial order parameter of the oscillators. In this way, we have observed a
variety of CS patterns in the two groups. Figs. \ref{Fig:breath}(a) and (b) show two typical CS patterns
after the transient process,
where (a) denotes the case of coupling strength $\lambda=\lambda'=1$ and the initial order parameter $R_1(0)=0.275$
and $R_2(0)=0.569$, and (b) denotes the case of coupling strength $\lambda=1.5$ and $\lambda'=1$ and the initial
order parameters $R_1(0)=0.1$ and $R_2(0)=0.569$. We see that in each case, one group is synchronized
with $R_2=1$ and the other has a different $R_1<1$, implying a breathing CS. In contrast, we numerically calculate
the theoretical Eq. (\ref{lowdimesion3}) and show the results in Figs. \ref{Fig:breath} (c) and (d), where the
difference between the initial phases of the two groups is taken as $\Delta\Phi=\Phi_2(0)-\Phi_1(0)=2\pi/3$.
In fact, Figs. \ref{Fig:breath} (c) and (d) can be considered as the corresponding theoretical results of
Figs. \ref{Fig:breath}(a) and (b). Comparing Fig. \ref{Fig:breath} (a) with (c) and Fig. \ref{Fig:breath}(b)
with (d), respectively, we see that the theoretical results are qualitatively consistent with the numerical
simulations.

\begin{figure}
\epsfig{figure=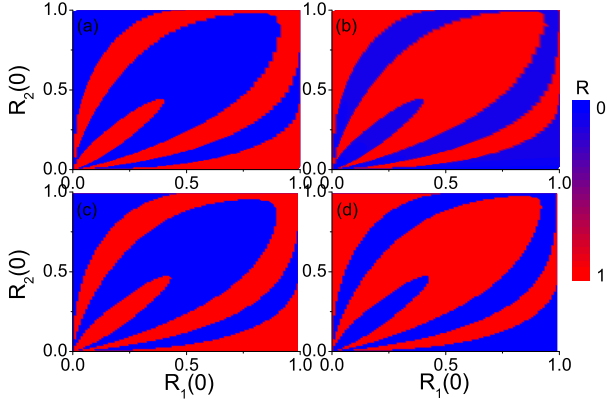,width=1.0\linewidth} \caption{(color
online). Influence of the initial order parameters $R_1(0)$ and $R_2(0)$ on the stabilized $R_1$ and $R_2$ in the
two groups. (a) and (b) show the results of numerical simulations for $R_1$ and $R_2$ from Eq. (\ref{model}),
respectively. (c) and (d)
show the theoretical results from Eq. (\ref{lowdimesion3}), corresponding to (a) and (b), respectively. The
difference between the initial average phases are all $\Delta\Phi=2\pi/3$ and the coupling strength is
$\lambda= \lambda'=1$. }
\label{Fig:initial}
\end{figure}

To show the dependence of CS on the initial conditions in details, we first fix the initial average phases as
$\Delta\Phi=\Phi_2(0)-\Phi_1(0)=2\pi/3$ and let the initial order parameters $R_1(0)$ and
$R_2(0)$ gradually increase from $0$ to $1$ by changing $x_0$ and $\eta$. Figs. \ref{Fig:initial}(a) and (b) show
how the stabilized $R_1$ and $R_2$ depend on the initial $R_1(0)$ and $R_2(0)$. Comparing Fig. \ref{Fig:initial}(a)
with (b) we see that $R_1$ is low when $R_2$ is high, and vice versa, i.e. they are complementary, indicating
that the whole system is always in CS. This is an interestingly finding which tells us that no matter what the
initial conditions are, we can
always find one group in high coherence while the other in low coherence, indicating that the basin of CS in the
model of Eq. (\ref{model}) is the whole initial condition space or CS is robust to initial conditions. This feature
is very different from some of the previous models of CS, where CS is typically observed for carefully chosen
initial conditions.
We also show the corresponding theoretical results from
Eq. (\ref{lowdimesion3}) in Figs. \ref{Fig:initial}(c) and (d). Comparing Fig. \ref{Fig:initial}(a) with (c) and
Fig. \ref{Fig:initial}(b) with (d), respectively, we see that they are almost the same, indicating the consistence
between the numerical simulations and theoretical results.

\begin{figure}
\epsfig{figure=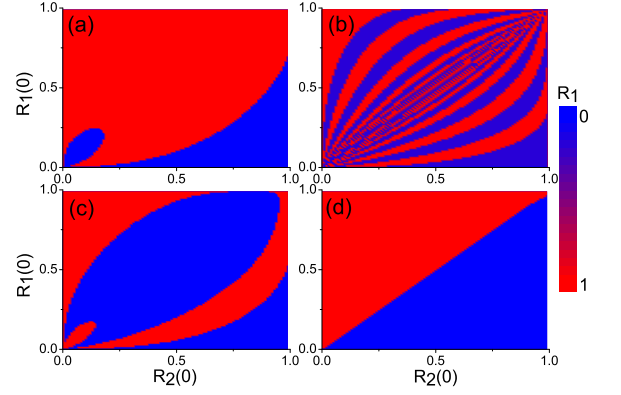,width=1.0\linewidth} \caption{(color
online). Influence of different initial average phases $\Delta\Phi=\Phi_2(0)-\Phi_1(0)$ on the patterns of CS
where the parameters are the same as in Fig. \ref{Fig:initial}. The results are obtained for the stationary $R_1$
by the theoretical analysis Eq. (\ref{lowdimesion3}). (a) Case of $\Delta\Phi=\pi/3$; (b) Case of $\Delta\Phi=\pi$;
(c) Case of $\Delta\Phi=\pi/2$; and (d) Case of $\Delta\Phi=0$.}
\label{Fig:initialphase}
\end{figure}

Then, we study the influence of the initial average phases. For this purpose, we consider a variety of difference
$\Delta\Phi=\Phi_2(0)-\Phi_1(0)$. As $\Delta\Phi$ is not neglected in Eq. (\ref{lowdimesion3}) of the dimensional
reduction, the low dimensional analysis shows the same effect with the numerical simulations by the circular
Cauchy distributed initial conditions. For this reason, we here only calculate the theoretical solution of
Eq. (\ref{lowdimesion3}). We find that the stabilized CS do depend on the specific value of the initial average phases.
As $R_1$ and $R_2$ are complementary, we here only calculate the stabilized $R_1$. Figure \ref{Fig:initialphase}
shows four typical cases where (a)-(d) represent the cases of $\Delta\Phi=\pi/3,\pi,\pi/2$ and $0$, respectively.
It is easy to see that the four patterns in Fig. \ref{Fig:initialphase} are different, indicating the diversity
of the CS basin patterns for different initial conditions.

\begin{figure}
\epsfig{figure=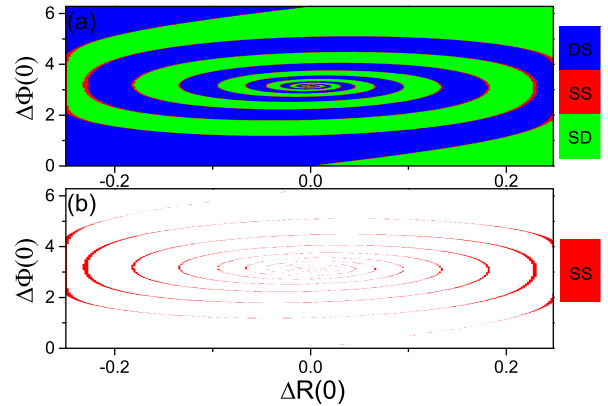,width=1.0\linewidth} \caption{(color
online). Influence of the initial order parameters, $\Delta R(0)$ and $\Delta\Phi(0)$, on the stabilized $R_1$ and
$R_2$ in the two groups with the same parameters as Fig. \ref{Fig:initial} and \ref{Fig:initialphase}, i.e.
$A=0$. The average of initial order parameter is $\langle R(0)\rangle=(R_1(0)+R_2(0))/2=0.75$.
Results are obtained by solving the Eq. (\ref{lowdimesion4}). (a) shows the distribution of the stabilized state
where DS means the first group is synchronized while the second group is desynchronized;
SD means the second group is synchronized while the first group is desynchronized; and SS means both of the two
groups are synchronized. (b) shows the basin of complete synchronized state only, where the basin of CS is hidden. }
\label{Fig:initial2}
\end{figure}

The robustness of CS to initial conditions is very interesting. To understand it better, we follow the Ref.
\cite{Abrams:2008} to make a further analysis on Eq. (\ref{lowdimesion3}). Firstly, we introduce a new parameter
$A=\lambda-\lambda'$. As all the frequencies of oscillators are zero, we rescale the coupling as $1=\lambda+\lambda'$
and thus obtain $\lambda=(1+A)/2$ and $\lambda'=(1-A)/2$. Therefore, $A=0$ represents the case of $\lambda=\lambda'$,
while $A=1$ represents the case of $\lambda'=0$, i.e. only one group of population. Then, we introduce
$\Delta\Phi=\Phi_2-\Phi_1$. For a
typical CS, one population is synchronized with $R=1$, thus we can set its order parameter as unity,
i.e. $R_1=1$ and $\dot{R_1}=0$. By this way, the Eq. (\ref{lowdimesion3}) becomes
\begin{eqnarray}
\label{lowdimesion4}
\dot{R}_2 &=& \frac{1}{2}R_2(1-R_2^2)[\frac{1+A}{2} R_2\cos{\alpha} \nonumber\\
& & +\frac{1-A}{2}\cos{(-\Delta\Phi+\alpha)}] \nonumber\\
\Delta\dot{\Phi} &=& \frac{1}{2}(1+R_2^2)[\frac{1+A}{2} R_2\sin{\alpha}+\frac{1-A}{2} \sin{(-\Delta\Phi+\alpha)}] \nonumber\\
& & -[\frac{1+A}{2}\sin{\alpha}+\frac{1-A}{2}R_2\sin{(\Delta\Phi+\alpha)}].
\end{eqnarray}
By letting $\dot{R_2}=0$, we can get three solutions: $R_2=1$, $R_2=0$ and
$R_2=-\frac{(1-A)\cos{(\alpha-\Delta\Phi)}}{(1+A)\cos{\alpha}}$. The first solution means a
complete synchronized state and the other two mean CS. By checking the values of $R_1$ and $R_2$ in both
Fig. \ref{Fig:initial} and Fig. \ref{Fig:initialphase}, we find that all the blue areas are in between $0.1$ and
$0.2$, indicating that they are the third solution. Therefore, we here focus only on the first two solutions, i.e.
$R_2=1$ and $R_2=0$. The Jacobian matrix of Eq. (\ref{lowdimesion4}) is
\begin{equation}\label{Jacobimatrix}
       M=\left[
         \begin{array}{cccc}
           a & b \\
           c & d \\
         \end{array}
         \right]
\end{equation}
with
\begin{eqnarray}
\label{jacobi}
a&=&\frac{1+A}{2}R_2\cos{\alpha}+\frac{1-A}{4}\cos{(\alpha-\Delta\Phi)} \nonumber\\
& & -(1+A)R_2^3\cos{\alpha}-\frac{3(1-A)}{4}R_2^2\cos{(\alpha-\Delta\Phi)} \nonumber\\
b&=&\frac{1}{2}R_2(1-R_2^2)\sin{(\alpha-\Delta\Phi)} \nonumber\\
c&=&\frac{1+A}{4}\sin{\alpha}+\frac{3(1+A)}{4}R_2^2\sin{\alpha} \nonumber\\
& & +\frac{1-A}{2}R_2\sin{(\alpha-\Delta\Phi)}-\frac{1-A}{2}\sin{(\Delta\Phi+\alpha)} \nonumber\\
d&=&-\frac{1-A}{4}(1+R_2^2)\cos{(\alpha-\Delta\Phi)} \nonumber\\
& & -\frac{1-A}{2}R_2\cos{(\Delta\Phi+\alpha)}.
\end{eqnarray}
By using of the linear stability analysis we find that the solution of $R_2=0$ is unstable while $R_2=1$ is stable with
the same parameters of Figs. \ref{Fig:initial} and \ref{Fig:initialphase}. It means that there is a probability
to observe the complete synchronization in the initial condition space. With further linear stability analysis,
we find that the invariant manifold with $R_1=R_2$ found in \cite{Martens:2016} still exists. In order to check this
point, we fix the
average of initial order $\langle R(0)\rangle=(R_1(0)+R_2(0))/2$ and look for the basin of the states in the plane
of $\Delta R(0)=R_1(0)-R_2(0)$ and $\Delta\Phi(0)$, i.e. taking the same way as Ref. \cite{Martens:2016} did.
Fig. \ref{Fig:initial2}(a) shows the distribution of the stabilized state for the case of $\langle R(0)\rangle=0.75$,
where DS means the first group is synchronized while the second one is desynchronized, SD means the second group
is synchronized while the first one is desynchronized, and SS means both of the two groups are synchronized.
Thus, DS and SD are CS while SS is a complete synchronized state. From
this figure, we see the basin of complete synchronized state (SS) is very narrow and it occurs only when the system
changes from DC state to SD state or the vice versa, which is the same as in Ref. \cite{Martens:2016}. To see it
more clear, Fig. \ref{Fig:initial2}(b) shows the basin of synchronized state only, where the basin of CS is hidden.
These basins of complete synchronized state are too narrow and thus make the SS state not easy to be observed, which
is the reason why we miss the complete synchronized state in Figs. \ref{Fig:initial} and \ref{Fig:initialphase}. On
the other hand, we notice that in Fig. \ref{Fig:initial2}, the basins of the states are spiral shaped around the
point $\Delta R(0)=0$, $\Delta\Phi(0)=\pi$, indicating the influence of the initial phases. This is very similar
with the result of Ref. \cite{Martens:2016}.

\begin{figure}
\epsfig{figure=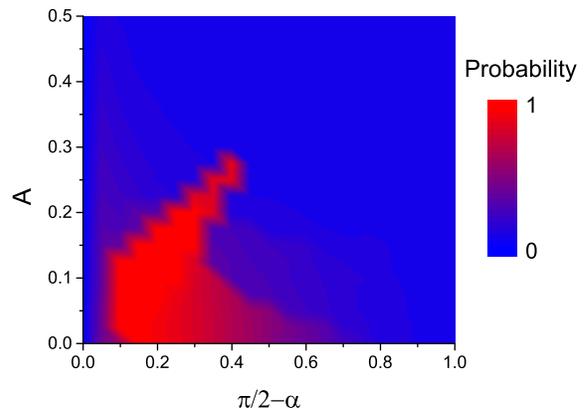,width=1.0\linewidth} \caption{(color
online). Probability of CS when using different initial conditions in the case of different $A$ and $\alpha$,
where the results are obtained by solving the Eq. (\ref{lowdimesion4}).}\label{Fig:possibility}
\end{figure}

In order to show how the basins of CS change with parameters, we calculate the probability of chimera state with
different initial conditions in the parameter plane of $A$ versus $\pi/2-\alpha$. Fig.\ref{Fig:possibility} shows the
results. It is easy to see that the probability of CS decreases with the decreasing of $\alpha$. When $A$ is large,
the probability of of CS becomes zero.

\section{Discussion}
To connect CS and ES, Eq. (\ref{model}) has three key aspects. The first one is the asymmetric couplings $\lambda$
and $\lambda'$. When $\lambda>\lambda'$, the coupling in each group is greater than that between the two
groups. Thus, the oscillators may be synchronized in their own groups but remain unsynchronized to those in
another group. The second one is the control parameter $\beta$. It guarantees
the appearance of ES. And the third one is the range parameter of the natural frequencies $\delta$. When $\delta$
is relatively large, we have both ES and CS-like behaviors. When $\delta$ is relatively small, we
only have CS. In this sense, we may also consider $\delta$ as the parameter connecting CS and ES.

One advantage of Eq. (\ref{model}) is that its CS can be easily observed. The underlying mechanism may be the
bistability. It is known that CS is a kind of symmetry breaking of coherence, due to the symmetry breaking in
initial conditions. If a system shows CS, its oscillators should have multi or bistability so that the
sensitivity to initial conditions can evolve into the final coexisting behaviors of coherence and incoherence
in different population groups. Thus, the multi or bistability is the necessary condition for CS. On the other
hand, a characteristic feature of ES is the existence of a hysteresis loop in the order parameter. When coupling
strength is located in this hysteresis region, the system has two stable states with one high coherence
and the other low coherence, separated by an unstable state. When the coupling is increased adiabatically in
the bistable region, the feature of low (high) coherence is remained and thus result in the hysteresis loop,
indicating that the bistability is also the necessary condition for ES. Therefore, the bistability is the common
basis of CS and ES.

Because of the correlation between local order parameter and coupling strength, our model is more sensitive to
symmetry breaking of initial conditions, which makes CS be observed easier. On the other hand, we find that in
our model, the basin of CS can be very large, which is similar with the large basin of CS in
Refs. \cite{Martens:2016,Feng:2015}. The spiral-shaped basin of states is also similar with the basin
structure in \cite{Martens:2016}, and reminds us the spiral wave chimeras in \cite{Martens:2009} although
they are different phenomena.

In conclusion, we have presented a model to describe both CS and ES. We reveal that in two limits, the system
goes to CS or ES, respectively. While in between the two limits, the model can show both CS and ES at the
same coupling strength. The frequency distribution parameter $\delta$ may seriously influence the final state.
When all the natural frequencies are zero, CS is robust to initial conditions and thus, a diversity of
the CS basin patterns can be observed. These findings have been confirmed by both numerical simulations and
theoretical analysis, which improves our understanding on both CS and ES, especially on their connection.

X.Z. thanks Prof. Arkady Pikvosky for many useful discussions. The authors thank the reviewers for their valuable
comments. This work was partially supported by the NNSF of China
under Grant Nos. 11135001 and 11375066, 973 Program under Grant No. 2013CB834100, and the Open Fund from the SKLPS of ECNU.

\end{document}